\documentclass[cits]{PoS}

\usepackage{graphics}
\usepackage{amsmath,amssymb}
\newcommand{\tr}{{\rm tr}\,}
\newcommand{\pmspec}{{\raisebox{-1.5pt}{\makebox[0pt][l]{\hspace{-3.57pt}{
          {\tiny[\,\,\,\,\, ]}}}}{\pm}}}

\title{Hadron-hadron total cross sections and soft high-energy 
  scattering on the lattice}

\ShortTitle{Hadron-hadron total cross sections and soft high-energy 
  scattering on the lattice}

\author{\speaker{Matteo Giordano}\thanks{Supported by MICINN under the
    CPAN project CSD2007-00042 from the Consolider-Ingenio2010
    program, and under the grant FPA2009-09638.}\\ 
        Departamento de F\'isica Te\'orica, Universidad de
        Zaragoza, \\ 
        Calle Pedro Cerbuna 12, E--50009 Zaragoza, Spain\\
        E-mail: \email{giordano@unizar.es}}
\author{Enrico Meggiolaro\\
        Dipartimento di Fisica, Universit\`a di Pisa, and INFN,
        Sezione di Pisa,\\
        Largo B.~Pontecorvo 3, I-56127 Pisa, Italy\\
        E-mail: \email{enrico.meggiolaro@df.unipi.it}}

\abstract{The nonperturbative approach to soft high--energy
  hadron--hadron scattering, based on the analytic continuation of
  Euclidean Wilson--loop correlation functions, makes possible the
  investigation of the problem of the asymptotic energy dependence of
  hadron--hadron total cross sections by means of lattice
  calculations. In this contribution we compare the lattice 
  numerical results to analytic results obtained with various
  nonperturbative techniques. We also discuss the possibility to
  obtain indications of the rise of hadron--hadron total cross
  sections with energy directly from the lattice data.}   

\FullConference{ The XXIX International Symposium on Lattice Field
  Theory - Lattice 2011\\ 
July 10-16, 2011\\
Squaw Valley, Lake Tahoe, California}

\begin{document}

\section{Introduction}

The problem of predicting total cross sections
at high energy from first principles is one of the oldest open
problems of hadronic physics (see, e.g.,~\cite{pomeron-book} and
references therein), not yet satisfactorily solved in QCD. This
problem is part of the more general problem of high--energy elastic
scattering at low transferred momentum, the so--called {\it soft
  high--energy scattering}. As soft high--energy processes possess two
different energy scales, the total center--of--mass energy squared  
$s$ and the transferred momentum squared $t$, smaller than the typical
energy scale of strong interactions ($|t| \lesssim 1~ {\rm GeV}^2 \ll
s$), we cannot fully rely on perturbation theory. A genuine
nonperturbative approach in the framework of QCD has been proposed  
in~\cite{Nachtmann91} and further developed in a number of papers
(see, e.g.,~\cite{pomeron-book} for a list of references): 
using a functional integral approach, high--energy hadron--hadron
elastic scattering amplitudes are shown to be governed by the
correlation function (CF) of certain Wilson loops defined in Minkowski
space. Moreover, as it has been shown
in~\cite{Meggiolaro97,Meggiolaro05,GM2006,GM2009}, such a CF can be
reconstructed by analytic continuation from the CF of two Euclidean 
Wilson loops, that can be calculated using the nonperturbative methods
of Euclidean Field Theory. 

In~\cite{GM2008,GM2010} we have investigated this problem by means of
numerical simulations in Lattice Gauge Theory (LGT). Although we
cannot obtain an analytic expression in this way, nevertheless this is
a first--principle approach that provides (within the errors) the true
QCD expectation for the relevant CF. In this contribution, after a
survey of the nonperturbative approach to soft high--energy scattering
in the case of meson--meson {\it elastic} scattering, we will present
our numerical approach based on LGT, compare the numerical results
to the existing analytic models, and discuss the possibility to obtain
indications of the rise of total cross sections directly from the
lattice data.

\section{High--energy meson--meson scattering and Wilson--loop
  correlation functions} 

We sketch here the nonperturbative approach to soft high--energy
scattering (see~\cite{GM2008} for a more detailed presentation). The
elastic scattering amplitudes of two mesons (taken for simplicity with
the same mass $m$) in the {\it soft} high--energy regime can be
reconstructed, after folding with the appropriate wave functions, from
the scattering amplitude ${\cal M}_{(dd)}$ of two dipoles of fixed
transverse sizes $\vec{R}_{1,2\perp}$, and fixed
longitudinal--momentum fractions $f_{1,2}$ of the two quarks in the
two dipoles~\cite{DFK}. In turn, the dipole--dipole ({\it dd})
scattering amplitude is obtained from the (properly normalised) CF of
two Wilson loops in the fundamental representation, defined in
Minkowski spacetime, running along the paths made up of the quark and
antiquark classical straight--line trajectories, and thus forming a
hyperbolic angle $\chi$ ($\simeq \log(s/m^2)$ at high energy). The
paths are cut at proper times $\pm T$ as an IR regularisation, and
closed by straight--line ``links'' in the transverse plane, in order
to ensure gauge invariance. Eventually, the limit $T\to\infty$ has to
be taken. 

It has been shown in~\cite{Meggiolaro97,Meggiolaro05,GM2006,GM2009}
that the relevant Wilson--loop CF can be reconstructed, by means of
analytic continuation, from the Euclidean CF of two Euclidean Wilson
loops, 
\begin{equation}
\label{GE}
{\cal G}_E(\theta;T;\vec{z}_\perp;1,2) \equiv
\dfrac{\langle \widetilde{\cal W}^{\,(T)}_1 \widetilde{\cal W}^{\,(T)}_2 \rangle_E}
{\langle \widetilde{\cal W}^{\,(T)}_1 \rangle_E
\langle \widetilde{\cal W}^{\,(T)}_2 \rangle_E } - 1\,, \,\,\,\,
\widetilde{\cal W}^{\,(T)}_{1,2} \equiv 
{\displaystyle\dfrac{1}{N_c}} \tr \left\{ {T}\!\exp
\left[ -ig \displaystyle\oint_{\widetilde{\cal C}_{1,2}}
 \tilde{A}_{\mu}(\tilde{x}) d\tilde{x}_{\mu} \right] \right\},
\end{equation}
where $\langle\ldots\rangle_E$ is the average in the sense of the
Euclidean QCD functional integral, and ``$1[2]$'' stand for
``$\vec{R}_{1[2]\perp}, f_{1[2]}$''. The Euclidean Wilson loops
$\widetilde{\cal W}^{\,(T)}_{1,2}$ are calculated on the following
straight--line paths,  
\begin{equation}
\widetilde{\cal C}_1 : \tilde{X}^{1q[\bar{q}]}(\tau) = z +
\frac{\tilde{p}_{1}}{m} \tau 
+ f^{q[\bar{q}]}_1 \tilde{R}_{1} , \quad
\widetilde{\cal C}_2 : \tilde{X}^{2q[\bar{q}]}(\tau) =
\frac{\tilde{p}_{2}}{m} \tau 
+ f^{q[\bar{q}]}_2 \tilde{R}_{2},
\label{trajE}
\end{equation}
with $\tau\in [-T,T]$, and closed by straight--line paths in the
transverse plane at $\tau=\pm T$. The four--vectors $\tilde{p}_{1}$
and $\tilde{p}_{2}$ are chosen to be $\tilde{p}_{1[2]}={m}(
\pmspec\sin\frac{\theta}{2}, \vec{0}_{\perp}, \cos\frac{\theta}{2})$
(taking $\tilde{X}_{4}$ to be the ``Euclidean time''), $\theta$ being
the angle formed by the two trajectories, i.e., $\tilde{p}_{1} \cdot
\tilde{p}_{2} = m^2 \cos\theta$. Moreover, $\tilde{R}_{i} =
(0,\vec{R}_{i\perp},0)$ and $\tilde{z} = (0,\vec{z}_{\perp},0)$. We
define also the CF with the IR cutoff removed as $\displaystyle {\cal
  C}_E \equiv \lim_{T\to\infty} {\cal G}_E $.  

The {\it dd} scattering amplitude is then obtained from ${\cal
  C}_E$ by means of analytic continuation as
\begin{equation}
{\cal M}_{(dd)} (s,t;1,2) 
\equiv -i~2s \displaystyle\int d^2 \vec{z}_\perp
e^{i \vec{q}_\perp \cdot \vec{z}_\perp}
{\cal C}_E(\theta\to -i\chi \mathop\sim_{s \to \infty} -i\log(s/m^2);
\vec{z}_\perp;1,2)\, ,
\label{scatt-loop}
\end{equation}
where $s \equiv (p_1 + p_2)^2$ and $t = -|\vec{q}_\perp|^2$ ($\vec{q}_\perp$
being the transferred momentum) are the usual Mandelstam variables
(for a detailed discussion on the analytic continuation
see~\cite{GM2009}, where  we have shown, on nonperturbative grounds,
that the required analyticity hypotheses are indeed satisfied). In the
following, without loss of generality~\cite{GM2010}, we will take the
longitudinal--momentum fractions $f_{1,2}=\frac{1}{2}$, and suppress
the dependence on $f_{1,2}$ in ${\cal G}_E$ and ${\cal C}_E$. 

\section{Wilson--loop correlation functions on the lattice}

The gauge--invariant Wilson--loop CF ${\cal G}_E$ is a natural
candidate for a lattice computation, but the explicit breaking of
$O(4)$ invariance on a lattice requires special care. As straight
lines on a lattice can be either parallel or orthogonal, we are forced
to use {\it off--axis} Wilson loops to cover a significantly large set
of angles~\cite{GM2008}. To stay as close as possible to the continuum
case, the loop sides are evaluated on the lattice paths that minimise
the distance from the continuum paths: this can be easily accomplished
by means of the well--known {\it Bresenham algorithm}~\cite{Bres}. The
relevant Wilson loops $\widetilde{\cal
  W}_L(\vec{l}_{\parallel};\vec{r}_{\perp};n)$ are then characterised
by the position $n$ of their center and by two 2D vectors
$\vec{l}_{\parallel}$ and $\vec{r}_{\perp}$, corresponding
respectively to the longitudinal and transverse sides of the loop. 
Setting $\widetilde{\cal W}_{L1} \equiv \widetilde{\cal
  W}_{L}(\vec{l}_{1\parallel};\vec{r}_{1\perp};d)$,
$d=(0,\vec{d}_{\perp},0)$, and
 $\widetilde{\cal W}_{L2} \equiv 
\widetilde{\cal W}_{L}(\vec{l}_{2\parallel};\vec{r}_{2\perp};0)$, we
%then 
define on the lattice
\begin{equation}
\label{eq:corr_lat}
{\cal G}_L(\vec{l}_{1\parallel},\vec{l}_{2\parallel};X_{\perp}) %&
\equiv
\frac{\langle \widetilde{\cal W}_{L1}
\widetilde{\cal W}_{L2} \rangle}
{\langle \widetilde{\cal W}_{L1}\rangle\langle \widetilde{\cal
    W}_{L2}\rangle} 
- 1, \quad
{\cal C}_L(\hat{l}_{1\parallel},\hat{l}_{2\parallel};X_\perp)
 \equiv \lim_{L_1,\,L_2\to\infty}
{\cal G}_L(\vec{l}_{1\parallel}, \vec{l}_{2\parallel};X_{\perp}).
\end{equation}
Here $X_\perp$ denotes collectively the relevant transverse variables,  
$X_\perp=(\vec{d}_{\perp};\vec{r}_{1\perp},\vec{r}_{2\perp})$. Also,
$\hat{l}_{i\parallel} \equiv \vec{l}_{i\parallel}/L_i$, where $L_i
\equiv |\vec{l}_{i\parallel}|$ are defined to be the lengths of the
longitudinal sides of the loops in lattice units. In the continuum
limit, where $O(4)$ invariance is restored, we expect 
\begin{equation}
\label{eq:contlim}
\begin{aligned}
&{\cal G}_L(\vec{l}_{1\parallel},\vec{l}_{2\parallel};X_\perp)
\mathop\simeq_{a\to 0}
{\cal  G}_E(\theta;T_1=\textstyle\frac{aL_1}{2},
T_2=\textstyle\frac{aL_2}{2};aX_\perp)
, &&
{\cal C}_L(\hat{l}_{1\parallel},\hat{l}_{2\parallel};X_\perp)
\mathop\simeq_{a\to 0}
{\cal C}_E(\theta;aX_\perp),
\end{aligned}
\end{equation}
where $\hat{l}_{1\parallel}\cdot\hat{l}_{2\parallel} \equiv
\cos\theta$ defines the relative angle $\theta$ and $a$ is the lattice spacing.
\begin{figure}[htb]
\centerline{\hfill
\raisebox{0.55cm}{\includegraphics[width=6.5 cm]{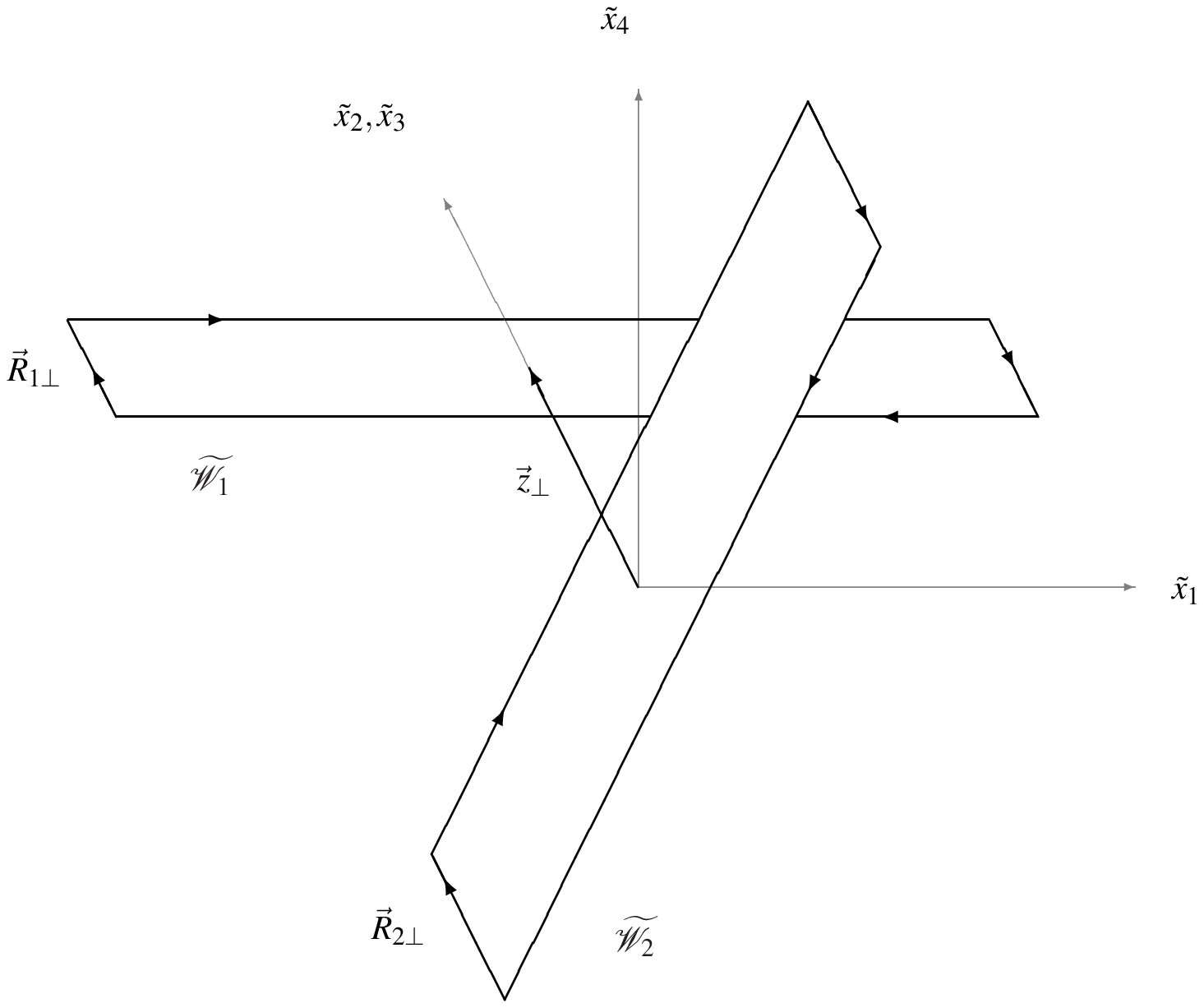}}
 \hfill
\includegraphics[width=5.3 cm]{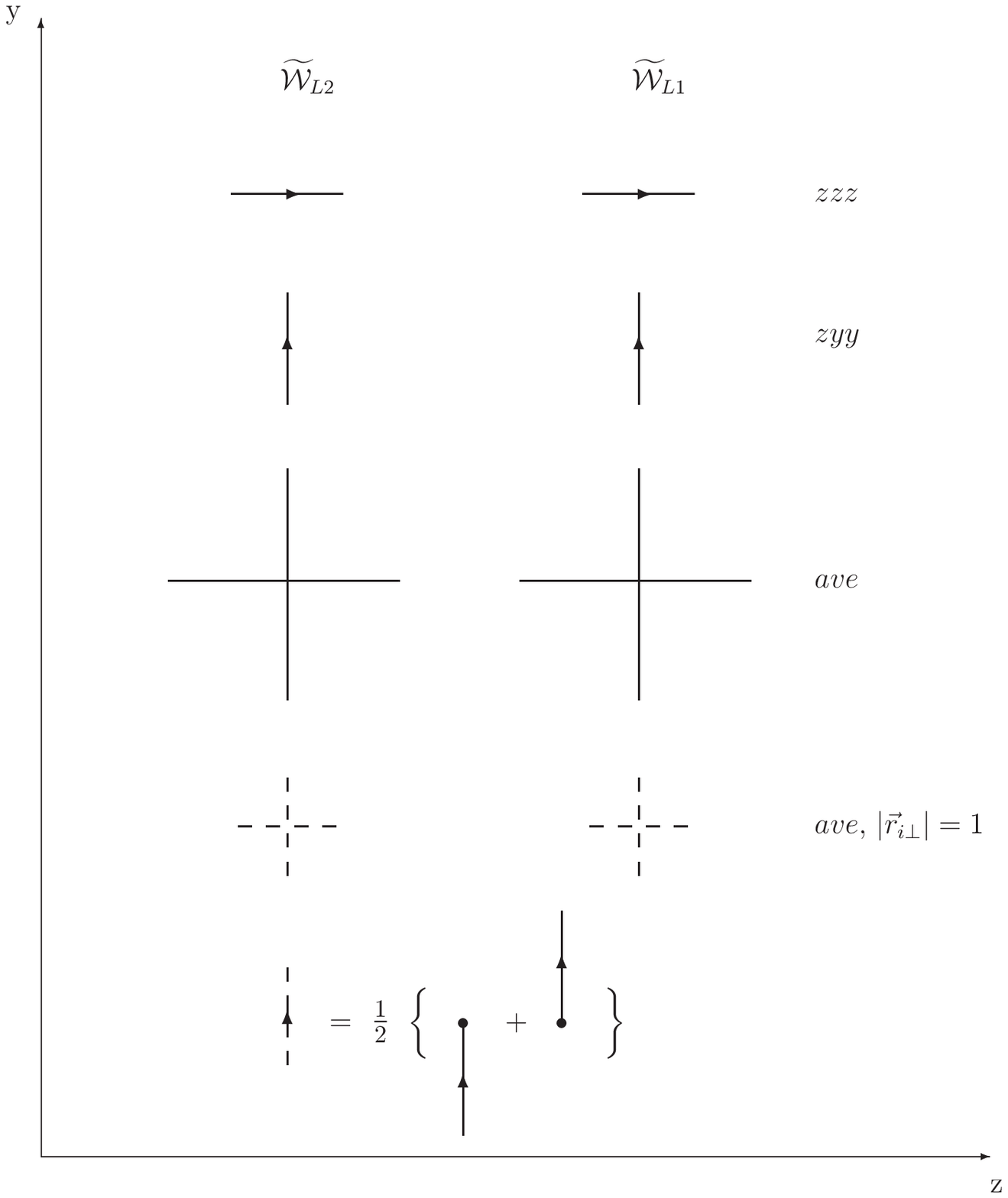}\hfill
}
\caption{(Left) The relevant Wilson--loop configuration. Using the $O(4)$
 invariance of the Euclidean theory we have put $\tilde{p}_{1}$ parallel to
 the $\tilde{x}_{1}$ axis. (Right) Loop configuration in the transverse
 plane.}
\label{fig:config}
\end{figure}

In~\cite{GM2008,GM2010} we have performed a Monte Carlo calculation of
${\cal G}_L$ for several values of the relative angle, various lengths
and different configurations in the transverse plane. We used 30000
{\it quenched} configurations generated with the $SU(3)$ Wilson action
at $\beta =6.0$, corresponding to $a\simeq 0.1\,{\rm fm}$, on a $16^4$
hypercubic lattice with periodic boundary conditions. This choice is
made in order to stay within the so--called ``{\it scaling window}'':
in this sense we are relying in an indirect way on the validity of the
relation (\ref{eq:contlim}) between Wilson--loop CFs on the lattice
and in the continuum. 

To keep the corrections due to $O(4)$ invariance breaking as small as
possible, we have kept one of the two loops {\it on--axis} and we have
only tilted the other one as shown in Fig.~\ref{fig:config}
(left). The on--axis loop $\widetilde{\cal W}_{L1}$ is taken to be
parallel to the $x_{E1}$ axis, $\vec{l}_{1\parallel}=(L_1,0)$, and of
length $L_1=6,8$, and we have used two sets of off--axis loops
$\widetilde{\cal W}_{L2}$ tilted at $\cot \theta = 0,\pm 1,\pm 2$. 
We have used loops with transverse size
$|\vec{r}_{1\perp}|=|\vec{r}_{2\perp}|=1$ in lattice units; the loop 
configurations in the transverse plane are those illustrated in
Fig.~\ref{fig:config} (right), namely 
$\vec{d}_{\perp} \parallel \vec{r}_{1\perp} \parallel
\vec{r}_{2\perp}$ (which we call ``{\it zzz}'') and $\vec{d}_{\perp} \perp
\vec{r}_{1\perp} \parallel \vec{r}_{2\perp}$ (``{\it zyy}''). We have
also measured the orientation--averaged quantity (``{\it ave}'')
defined as ${\cal
  C}_E^{ave}(\theta;\vec{z}_{\perp};|\vec{R}_{1\perp}|,|\vec{R}_{2\perp}|)  
\equiv \int d\hat{R}_{1\perp} \int d\hat{R}_{2\perp}
{\cal C}_E(\theta;\vec{z}_{\perp};\vec{R}_{1\perp},\vec{R}_{2\perp})$, 
where $\int d\hat{R}_{i\perp}$ stands for integration over the orientations of
$\vec{R}_{i\perp}$. The lattice version of this equation is easily
recovered for even (integer) values of the transverse sizes; in our
particular case, $|\vec{r}_{i\perp}|=1$, we have to use a sort of
``{\it smearing}'' procedure, averaging nearby loops as depicted in
Fig.~\ref{fig:config} (right).

Since we are interested in the limit $T\to\infty$, we have to perform
it on the lattice by looking for a {\it plateau} of ${\cal G}_L$
plotted against the loop lengths $L_{1,2}$. On a $16^4$ lattice it is
difficult to have a sufficiently long loop while at the same time
avoiding finite size effects, and at best we can push the calculation
up to $L=8$. Nevertheless, our data show that ${\cal G}_L$ is already
quite stable against variations of the loop lengths at $L_{1,2}\simeq
8$ (at least for $\theta$ not too close to $0^{\circ}$ or
$180^{\circ}$, where it is expected to diverge due to its relation
with the static {\it dd} potential, see~\cite{GM2008,GM2010}) 
and so we can take the data for the largest loops available as a
reasonable approximation of ${\cal C}_L$.

We have considered the values $d=0,1,2$ for the distance between the
centers of the loops: as expected, the CFs vanish rapidly as $d$
increases, thus making the calculation with our simple  ``brute
force'' approach very difficult at larger distances. 

\section{Comparison with analytical results and the problem of 
   total cross sections}

As already pointed out in the Introduction, numerical simulations of
LGT can provide the Euclidean CF only for a finite set of
$\theta$--values, and so its analytic properties cannot be directly
attained; nevertheless, they are first--principles calculations that
give us (within the errors) the true QCD expectation for this
quantity. Approximate analytic calculations of this same CF have then
to be compared with the lattice data, in order to test the goodness of
the approximations involved. This can be done either by direct
comparison, when a numerical prediction is available, or by fitting
the lattice data with the functional form provided by a given
model. The Euclidean CFs we are interested in have been evaluated in
the {\it Stochastic Vacuum Model} (SVM)~\cite{LLCM2}, in the {\it
  Instanton Liquid Model} (ILM)~\cite{instanton1,GM2010}, and using
the AdS/CFT correspondence~\cite{JP1}: the comparison of our data with
these analytic calculations is not, generally speaking, fully
satisfactory. 

\begin{figure}[t]
\centerline{
\hfill\includegraphics[width=5.9 cm]{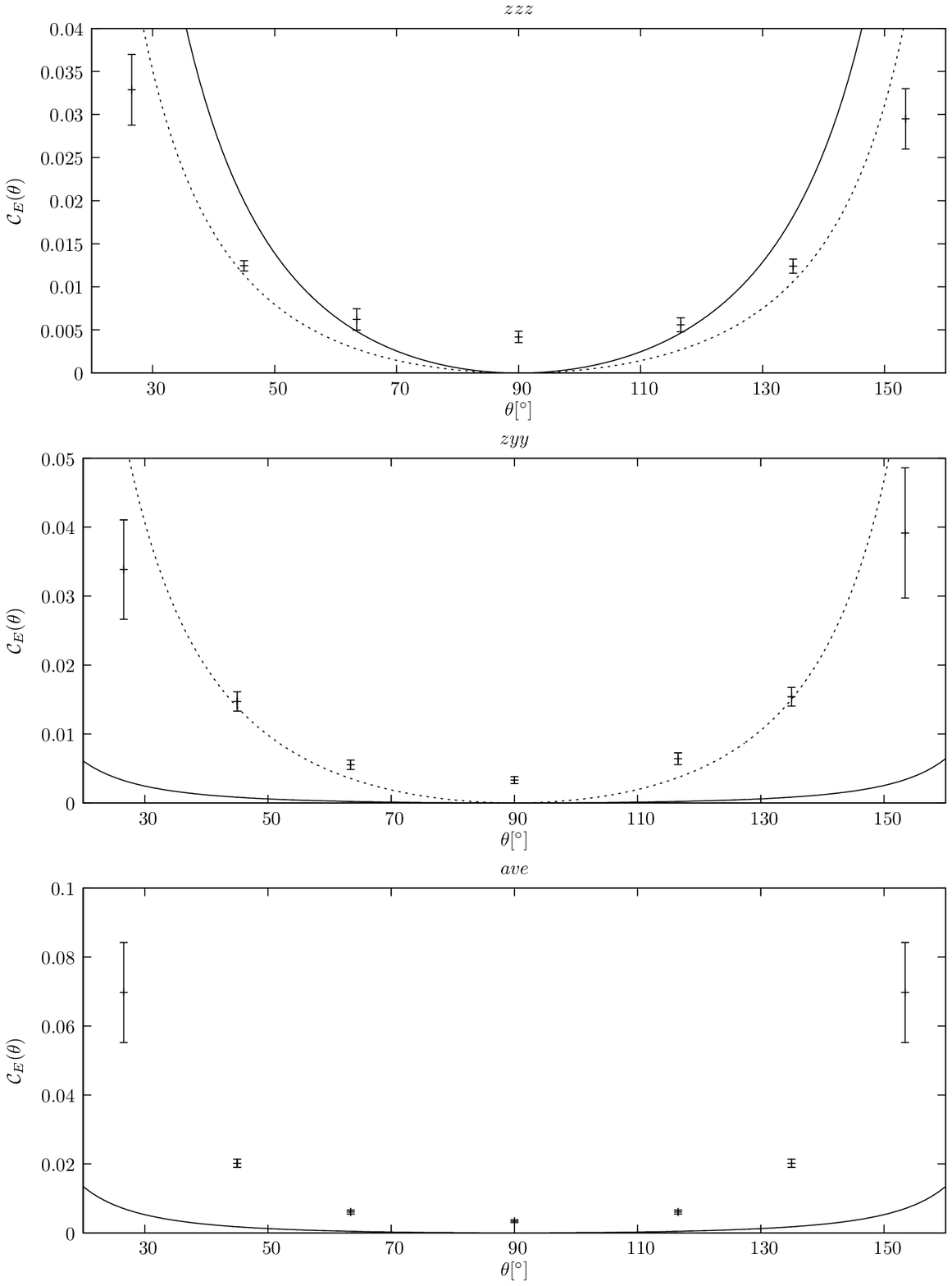}\hfill\includegraphics[width=5.9 cm]{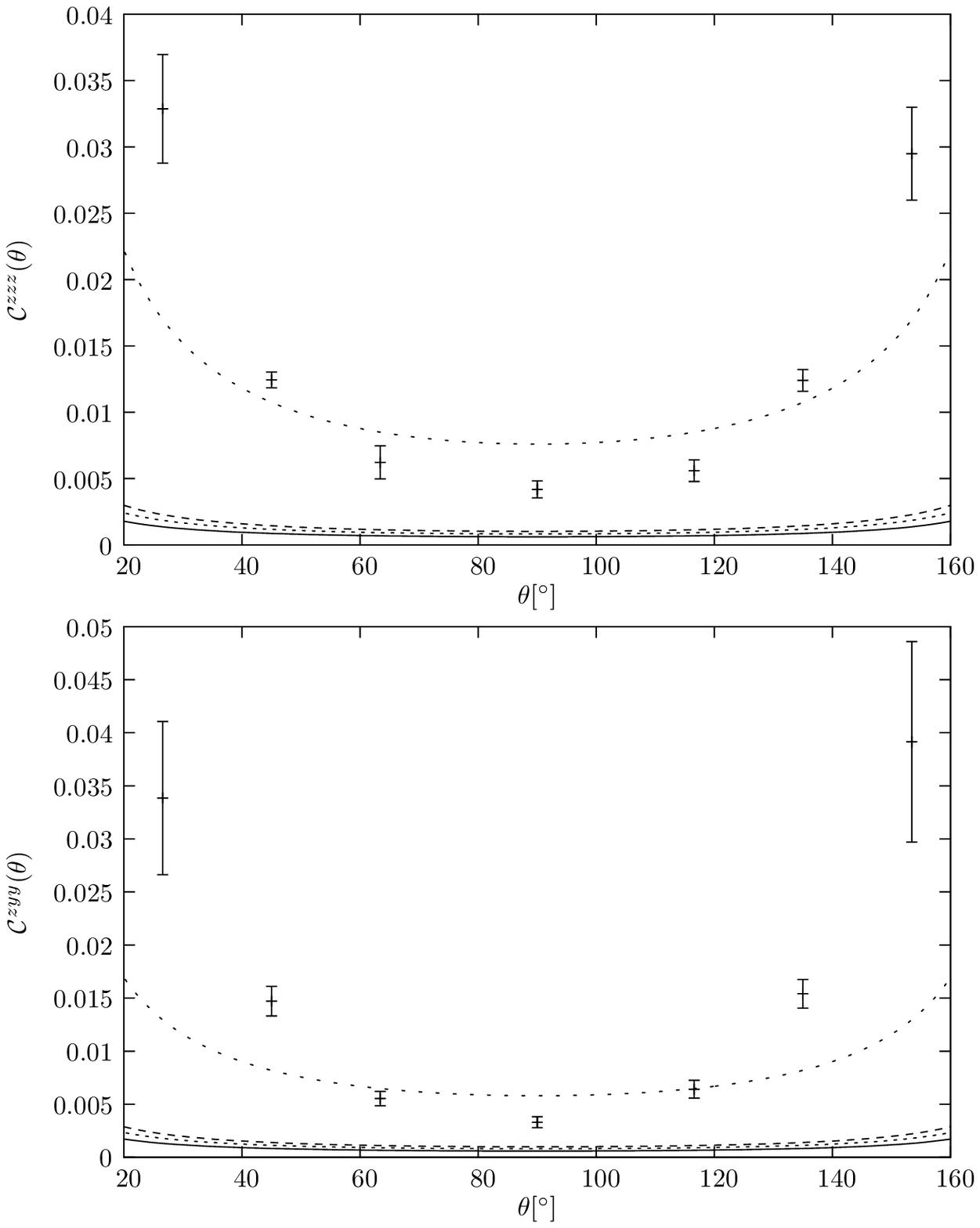}\hfill
}
\caption{(Left) Comparison of the lattice data to the SVM
  expression~\protect\eqref{eq:SVM}, with $K_{\rm SVM}$ calculated
  according to~\cite{LLCM2} (solid line), or determined through a
  best--fit (for the ``{\it zzz}'' and ``{\it zyy}'' cases only,
  dotted line) at $d=1$. (Right) Comparison of the lattice data to the
  ILM expression~\protect\eqref{eq:instanton} with $K_{\rm ILM}$
  calculated according to~\cite{GM2010} (dotted, dashed, and solid
  line, corresponding to different values of the model parameters), or
  determined through a best--fit (sparse dotted line) for the ``{\it
    zzz}'' and ``{\it zyy}'' cases at $d=1$.}  
\label{fig:svm-ilm}
\end{figure}
In the SVM~\cite{LLCM2} the Wilson--loop CF is given by the expression
\begin{equation}
\label{eq:SVM}
{\cal C}^{\,\rm (SVM)}_E(\theta) = \frac{2}{3}
\exp\left(-\frac{1}{3} K_{\rm SVM}\cot\theta\right)
+ \frac{1}{3} \exp\left(\frac{2}{3} K_{\rm SVM}\cot\theta\right) - 1\,,
\end{equation}
where $K_{\rm SVM}$ is a function of $\vec{z}_{\perp}$, $\vec{R}_{1\perp}$
and $\vec{R}_{2\perp}$ only, given in~\cite{LLCM2}, that we have used
to evaluate \eqref{eq:SVM} numerically in the relevant cases. The SVM
prediction \eqref{eq:SVM} agrees with our lattice data in a few cases,
at least in the shape and in the order of magnitude, but, in general, 
it is far from being satisfactory, see Fig.~\ref{fig:svm-ilm} (left). 
The same conclusion is reached if one performs instead a
one--parameter ($K_{\rm SVM}$) best--fit with the given expression:
the values of the chi--squared per degree of freedom ($\chi^2_{\rm
  d.o.f.}$) of this and the other fits that we have performed are
listed in Table \ref{tab:chi2}.

The ILM predicts the following functional form of the
CF~\cite{instanton1,GM2010}, 
\begin{equation}
\label{eq:instanton}
  {\cal C}^{\,\rm (ILM)}_E(\theta) = \frac{K_{\rm ILM}}{\sin\theta}\,;
\end{equation}
in particular, a well--defined numerical {\it prediction} for $K_{\rm
  ILM}$ has been obtained in~\cite{GM2010}. The ILM prediction turns
out to be more or less of the correct order of magnitude in the range
of distances considered, at least around $\theta=\frac{\pi}{2}$, but
it does not match properly the lattice data; the same mismatch is seen
also in a best--fit with Eq.~\eqref{eq:instanton}, see
Fig.~\ref{fig:svm-ilm} (right). Moreover, the ILM prediction seems to
overestimate the correlation length which sets the scale for the rapid
decrease of the CF with the distance between the loops: this is also
supported by the comparison of the prediction for the
instanton--induced {{\it dd} potential} $V_{dd}$ with some preliminary
numerical results on the lattice~\cite{GM2010}. It is worth noting
that largely improved best--fits (see Table \ref{tab:chi2}) are
obtained by combining Eq.~\eqref{eq:instanton} with the functional
form  ${\cal C}_E^{(\rm pert)}(\theta) = K_{\rm pert} (\cot\theta)^2$,
corresponding to the leading--order result in perturbation
theory~\cite{BB,Meggiolaro05,LLCM2}, into the following expression,
${\cal C}^{\,\rm (ILMp)}_E(\theta) = \frac{K_{\rm ILMp}}{\sin\theta}
+ K_{\rm ILMp}'(\cot\theta)^2$.

\begin{table}[t]
\centering
{\footnotesize 
 \begin{tabular}[h]{l|cc|ccc|ccc}
\hline
\hline
$\chi^2_{\rm d.o.f.}$ & \multicolumn{2}{c|}{$d=0$} &
\multicolumn{3}{c|}{$d=1$} & \multicolumn{3}{c}{$d=2$}\\
& {\it zzz}/{\it zyy} & {\it ave} & {\it zzz} & {\it zyy} &
{\it ave} & {\it zzz} & {\it zyy} & {\it ave} \\
\hline
SVM & 51 & - & 16 & 12 & - & 1.5 & 2.2 & -\\
pert & 53 & 34 & 16 & 13 & 13 & 1.5 & 2.2 & 4.5\\
ILM & 114 & 94 & 14 & 15 & 45 & 0.45 & 0.35 & 1.45\\
ILMp & 20 & 9.4 & 0.54 & 0.92 & 1.8 & 0.13 & 0.12 & 0.19\\
AdS/CFT & 40 & - & 1 & 0.63 & - & 0.14 & 0.065 & -\\
\hline
\end{tabular}}
\caption{Chi--squared per degree of freedom for a best--fit with the
indicated function.}
\label{tab:chi2}
\end{table}

Finally, we have tried a best--fit with the expression obtained
through AdS/CFT for the ${\cal N}=4$ SYM theory at large $N_c$, large
't Hooft coupling and large distances between the loops~\cite{JP1}: 
\begin{equation}
\label{eq:AdS}
{\cal C}^{\,\rm (AdS/CFT)}_E(\theta)
= \exp\left\{\dfrac{K_1}{\sin\theta} + K_2\cot\theta +
K_3\cos\theta\cot\theta\right\}-1\,.
\end{equation} 
Taking into account that it is a three--parameter best--fit, even this one
is not satisfactory: best--fits with QCD--inspired expressions with
only two parameters, like, e.g., the ILMp expression [or some
appropriate modification of the SVM expression \eqref{eq:SVM}] give
smaller $\chi^2_{\rm d.o.f.}$ (see Table \ref{tab:chi2}). 

As an important side remark, we note that our data show a clear signal
of $C$--odd contributions in {\it dd} scattering, which are related
through the {\it crossing--symmetry relations}~\cite{GM2006} to the 
antisymmetric part of ${\cal C}_E(\theta)$ with respect to
$\theta=\frac{\pi}{2}$. An asymmetry is present in the ``$zzz$'' and
``$zyy$'' tranverse configurations (${\cal C}_E^{ave}$ is trivially
symmetric), thus signalling the presence of {\it odderon}
contributions to the {\it dd} scattering amplitudes. Although these
$C$--odd contributions are averaged to zero in meson--meson scattering
(at least in our model), they might play a non--trivial role in
hadron--hadron processes in which baryons and antibaryons are also
involved. 

As we have said in the Introduction, the main motivation in studying
soft high--energy scattering is that it can lead to a resolution of
the total cross section puzzle. From this point of view, a
satisfactory comparison of the lattice data with the SVM or the ILM
would not have helped, since they yield constant or vanishing cross
sections at high energy, as it can be seen by using
Eqs.~\eqref{scatt-loop} and the {\it optical theorem}. An ambitious
question that one can ask at this point is if the lattice data are
compatible with rising total cross sections. An answer can in
principle be obtained by performing best--fits to the lattice data
with more general functions, leading to a non--trivial dependence on
energy. This approach requires special care, because of the analytic
continuation necessary to obtain the physical amplitude from the
Euclidean CF: one has therefore to restrict the set of admissible
fitting functions by imposing physical constraints (e.g., unitarity).  

In this framework, a possible strategy is suggested by the improvement
of best--fits achieved with the ILMp expression: the idea is to combine
known QCD results and variations thereof. As an example, one could
consider exponentiating two--gluon exchange and the one--instanton
contribution (i.e., the ILMp expression), and supplementing it with a
term which could yield a rising cross section, e.g., ${\cal C}_E^{\rm
  (rise)} = \exp\{{\cal C}_E^{\rm (ILMp)}\}\exp\{A_{\rm
  rise}(b)\left(\frac{\pi}{2}-\theta\right)^4(\cot\theta)^2\}-1$.  
Such an expression yields an amplitude respecting unitarity if $A_{\rm
  rise}(b)>0$ for large $b=|\vec{z}_\perp|$, and leads to the limit
behaviour $\sim (\log s)^2$ allowed by the Froissart bound if $A_{\rm
  rise}(b)\sim b^{-4}$ for large $b$. 

Another possible strategy is suggested by the AdS/CFT expression
\eqref{eq:AdS}: one can try to adapt to the case of QCD the analytic
expressions obtained in related models, such as ${\cal N}=4$ SYM.
As it has been shown in~\cite{GP2010}, by combining the knowledge of
the various coefficient functions $K_i$ in \eqref{eq:AdS} at large
$b$~\cite{JP1} with the unitarity constraint in the small--$b$
region, a non--trivial high--energy behaviour for the {\it dd} total cross
section in ${\cal N}=4$ SYM can emerge (including a {\it
  pomeron}--like behaviour $\sigma \sim s^{1/3}$). 
Although of course Eq.~\eqref{eq:AdS} is not expected to describe QCD,
it is sensible to assume in this case a similar functional form
(basically assuming the existence of the yet unknown gravity dual for
QCD). Assuming moreover that the known power--law behaviour of the $K_i$'s
(expected for a conformal theory) goes over into an exponentially
damped one (expected for a confining theory), in particular $K_3\to c
e^{-\mu b}$, one obtains a rising total cross section proportional to
the limit behaviour $\sim (\log s)^2$.  

It seems then worth investigating further the dependence of the
CFs on the relative distance between the loops, as well as on the
dipole sizes, as they could combine non--trivially with the dependence
on the relative angle: these and other related issues (including the
above--mentioned more general best--fits) are currently
under study~\cite{GMM}, and will be addressed in future works.

\end{document}